# On the Need for a Classification System for Consistent Characterization of the Composition of Planetary Bodies

David G. Russell

Owego Free Academy, 1 Sheldon Guile Blvd, Owego, NY USA

Email: russelld@oacsd.org

**Abstract**

A classification system is presented for characterizing the composition of planetary bodies. From mass-radius and mass-density relationships, planets may be broadly grouped into five composition classes identified as: Gas Giant, Rock-Ice Giant, gas-rich Terrestrial, Rock Terrestrial, and Rock-Ice Terrestrial based upon the mass fractions of H-He gas, rock, and ice. For each of these broad composition classes, specific bulk composition classes are defined and characterized with Solar System analog names. The classification system allows for both general and detailed characterization of exoplanets based upon planetary mass-radius-composition models and provides rationale for distinguishing "gas-rich super-Earths" from "mini-Neptunes".

## 1 Introduction

To date over 5000 exoplanets have been confirmed (NASA Exoplanet Archive – Akeson et al. 2013). One important goal of exoplanetary research is to characterize the composition and structure of these exoplanets. For planets with accurate mass and radius measurements it is possible to determine the bulk density and, using mass-radius-composition models, describe plausible bulk compositions (e.g. Grasset et al. 2009; Lopez&Fortney 2014; Zeng et al. 2016, 2019; Aguichine et al. 2021). In addition, laboratory studies have improved the equation of state (EoS) models used for developing interior structure models (Zeng & Sasselov 2013; Hakim et al. 2018a; Helled 2018; Miozzi et al. 2018; Helled et al. 2019; Helled & Fortney 2020; Huang et al. 2021; Vazan et al. 2022) and estimating the core mass fraction (Zeng et al. 2016). Since planetary composition and structure are affected by the composition of planetary building blocks, studies of stellar abundance ratios provide valuable information for refining models of bulk composition and interior structure (Fortney 2012; Santos et al. 2017; Dorn et al. 2015, 2017; Wang et al. 2018, 2019; Bitsch & Battistini 2020; Adibekyan et al. 2021).

Despite improvement in data, characterizing planetary composition and structure is still a problem that suffers from model degeneracy and model assumptions (Dorn et al. 2017). While degeneracy is most significant in the "Neptune" and "sub-Neptune" class of planets (Wolfgang & Lopez 2015; Lozovsky et al. 2018; Jontof-Hutter 2019; Bean et al. 2021; Vazan et al. 2022), degeneracy problems persist in characterizing the mineralogy of "Earth-like" terrestrial planets (Dorn et al. 2015; Hakim et al. 2018a).

Analysis of the mass-radius (M-R) and mass-density (M-$\rho$) relationships for the exoplanet population indicates the existence of several break points that divide the planetary population into mass ranges with a narrower set of planetary compositions (Hatzes & Rauer 2015; Chen & Kipping 2017; Bashi et al. 2017; Fulton et al. 2017; Otegi et al. 2020). M-R and M-$\rho$ relationships indicate that the population of planets with structure dominated by gravitational self-compression of H and He gas (gas giants) begins at approximately 120 $M_\oplus$ and extends to roughly 60 Jupiter masses ($M_J$) (Hatzes & Rauer 2015; Chen & Kipping 2017; Bashi et al. 2017). Planets with masses from 5 to 120 $M_\oplus$ include a mix of lower mass gas giants, Neptune-like planets with a rock or rock-ice interior and large H-He envelope, terrestrial planets, and rock or rock-ice composition planets with a small percentage by mass H-He envelope (Lopez & Fortney 2014; Chen & Kipping 2017; Zeng et al. 2019; Otegi et al. 2020; Owen et al. 2020), or alternatively a massive steam envelope (Turbet et al. 2020; Mousis et al. 2020; Aguichine et al. 2021). Planets with mass less than approximately 5 $M_\oplus$ are less likely to both accrete and maintain a significant H-He envelope and will more often have a rock or rock-ice composition (Otegi et al. 2020).

The break points identified from M-R and M-$\rho$ relationships help to narrow the range of composition types within each mass range, but an overlap of composition types remains. For example, the rocky planet population appears to overlap the "Neptune" planet range from about 5 to 26 $M_\oplus$ (Otegi et al. 2020) whereas the gas giant and "Neptune" populations overlap from 50 to 100 $M_\oplus$ (Petigura et al. 2017; Otegi et al. 2020; Millholland et al. 2020).

While it appears that all planets and brown dwarfs from approximately 0.4 to 60 $M_{Jup}$ have an H-He dominated composition, the primary formation mechanism may shift at different mass points within the mass range (Ma & Ge 2014; Santos et al. 2017; Narang et al. 2018; Schlaufman 2018; Goda & Matsuo 2019). Schlaufman (2018) found that core accretion is likely to be the primary formation mechanism for gas giants with masses less than 4 $M_{Jup}$ whereas disk instability appears to be a more prevalent formation mechanism at masses greater than 10 $M_{Jup}$. However, Adibekayn (2019) did not find evidence for a break point in formation mechanisms at 4$M_{Jup}$ instead finding that environmental conditions (disk mass, metallicity) may play a larger role in formation channels. In addition, brown dwarfs, formed by star-like gas collapse (Kumar 1963), may have masses as small as 4 to 5 $M_{Jup}$ (Caballero 2018; Luhman & Hapich 2020) but determining the specific formation mechanism for individual super-Jupiter mass gas giants and brown dwarfs remains problematic (Chabrier et al. 2014).

The Terrestrial planet class has a large range of possible compositions including silicate dominated (Earth), metal dominated (Mercury), carbon enriched (Bond et al. 2010; Hakim et. al. 2018; Miozzi et al. 2018; Hakim et al. 2019a,b; Allen-Sutter et al. 2020), and "ocean worlds" with varying fractions of rock and ice (e.g. Europa vs. Ganymede; Grasset et al. 2009; Jontof-Hutter 2019; Bean et al. 2021). In addition, the specific mineralogy of individual terrestrial planets may vary depending upon stellar abundance ratios (Dorn et al. 2015, 2017; Santos et al. 2017; Wang et al. 2018; Adibekyan et al. 2021).

Given the large variety of planetary compositions and formation mechanisms, the degeneracy in determining interior structure and bulk composition, and overlapping mass ranges for different composition classes, characterizing exoplanet composition classes in a consistent way has been challenging. For example, Hatzes & Rauer (2015) designated planets with masses from 0.3 to 60 $M_{Jup}$ as "giant planets" and planets with masses < 0.3 $M_{Jup}$ as "low mass planets" whereas Bashi et al. (2017) used the terms "small planets" and "large planets" to describe the same two populations. Goda & Matsuo (2019) suggested the giant planet population from 0.3 to 60 $M_J$ should be divided into three mass ranges identified as "intermediate-mass planets" (0.3-4 $M_{Jup}$), "massive planets" (4-25 $M_{Jup}$) and "brown dwarfs" (>25 $M_{Jup}$). Chen & Kipping (2017) identified the giant planets as "Jovian" worlds and split the low mass planets into two groups: "Neptunian worlds" and "Terran worlds". Zeng et al. (2019) used four radius ranges to divide the planetary population into "Rocky worlds", "Water worlds", "Transitional planets", and "Gas giants".

While the categorization of exoplanets is varied in the literature, the terminology suggested by Chen & Kipping (2017) closely aligns with the three broad classes of planets found in the solar system: gas giants (Jovian), ice giants (Neptunian), and terrestrial (Terran). However, the planetary population within the "Neptunian" composition mass range also includes "Terran" and "Jovian" planets (see samples and analysis in Otegi et al. 2020; Millholland et al. 2020). In addition, while Neptune-like planets are often characterized as "ice giants" it is possible that Neptune and Uranus could be rock dominated by mass, rather than ice dominated, or a "rock giant" composition (Helled et al. 2019; Helled & Fortney 2020; Teanby et al. 2020). Some Neptunian worlds may consist of a nearly pure rock body with just a small percentage by mass H-He envelope (Lopez & Fortney 2014; Zeng et al. 2019; Owen et al. 2020; Millholland et al. 2020; Bean et al. 2021) – a class of planet not found in the Solar System.

Complicating the characterization and communication of exoplanet discoveries is the use of terms such as "super-Earth", "super-Mercury", "super-Ganymede", "sub-Neptune", "super-Neptune", and "sub-Saturn" that are applied to exoplanets which often have a mass, radius, structure, composition, or stellar flux significantly different from the Solar System bodies they are compared with. The designation of a planet as a "sub-" or "super-" Earth, Neptune, or Saturn is further complicated by the fact that these designations may be applied based upon mass, radius, or composition – each of which can lead to a different class for the same planet. For example, a recent analysis of "sub-Saturns", identified as planets with radius from 4.0 to 8.0 $R_⊕$, included planets with mass ranging from 3.9 $M_⊕$ to 59.4 $M_⊕$ (Millholland et al. 2020 – hereafter MPB20). In terms of radius, planets in the MPB20 sample are "sub-Saturns". However, in terms of mass, the sample ranges from "super-Earth" to "sub-Saturn mass" with roughly half the sample in a "Neptune" mass range of 10-30 $M_⊕$. The envelope fractions determined in the MPB20 analysis revealed that these "sub-Saturn" radius planets can be characterized, by mass and composition, as "sub-Neptunes", "Neptunes", and "super-Neptunes", but with no actual "mini-gas giants" or "Saturn" composition planets. As an extreme example, the planet Kepler 87c is a super-Earth mass (6.4 +/- 0.8 $M_⊕$), sub-Saturn radius planet (6.14 +/- 0.29 $R_⊕$), with a Neptune composition MPB20.

The discussion above highlights the need for a consistent classification system for characterizing and communicating the composition of planetary bodies. An example of such a composition classification system is presented in this paper. The system includes familiar terminology for five broad composition classes and Solar System analog names for characterizing possible bulk compositions within each of the broad composition classes. The goal is to provide a classification system that is flexible enough to serve as a framework for communicating planetary science discoveries to professionals, students, and the public without becoming outdated as new discoveries are made. The paper is organized as follows: General definitions related to planet classification and composition are discussed in section 2. The five broad composition classes are described in section 3. Mass ranges for the composition classes are discussed in section 4. Solar System analog names for describing bulk composition classes are described in section 5. Section 6 is the discussion and section 7 provides a conclusion.

## 2 Definitions

### ~2.1 *Planetary composition components*

Three types of materials provide the building blocks for planetary bodies and are identified as "rock", "ice", and "gas" (Stern & Levison 2002; dePater & Lissauer 2015). Depending upon planetary mass, formation history, and post formation evolution, individual planets may have significant mass fractions of one, two, or all three of these composition components:

**Rock**: Rock is primarily composed of the elements Mg, Si, O, and Fe (Baraffe et al. 2014). Pure rock planets will frequently have silicate minerals differentiated into a crust and mantle with an iron-rich metal alloy core. Silicate mineralogy and iron core mass fraction can vary with the abundance of elements such as Ca, Mg, Na, Al, Ni, and Fe (Dorn et al. 2017; Unterborn et al. 2017; Santos et al. 2017; Adibekyan et al. 2021). Some rock planets may be enriched in carbon and have a graphite crust (Bond et al. 2010; Hakim et al. 2019a,b).

**Ice**: The elements carbon, nitrogen, hydrogen, oxygen, and sulfur form molecules such as $H_2O$, $H_2S$, $CH_4$, $NH_3$, $CO$, $N_2$, and $CO_2$ that are collectively referred to as 'astrophysical' or 'planetary' ices. (Stern & Levison 2002; Baraffe et al. 2014; de Pater & Lissauer 2015). Due to planetary formation mechanisms (see review by Raymond et al. 2018), pure ice composition planets should be rare. Most small ice-rich Dwarf planets and Satellite planets will have a rock composition core with overlying ice and interior ocean layers (dePater & Lissauer 2015). For planets >~1 Earth mass rock and ice may have >99% miscibility in the interior and therefore fully mixed rock-ice interiors with density gradients are possible (Vazan et al. 2022).

**Gas**: The elements H and He are "gas" and are the main mass component of stars, brown dwarfs, and gas giant planets (Baraffe et al. 2010).

It is important to emphasize that the terms "rock", "ice", and "gas", as described above, indicate composition only and do not require that the rock, ice, or gas exist in a specific state (solid, liquid, gas). Planetary interior structure models indicate that rock, ice, and gas can exist in multiple states within the same planet (e.g. dePater & Lissauer 2015) and models for different planet compositions are continuously revised as data and EoS models improve (Wahl et al. 2017; Hakim et al. 2018a, 2019a; Miozzi et al. 2018; Mazevet et al. 2019; Turbet et al. 2020; Vazan et al. 2022; Aguichine et al. 2021; Huang et al. 2021).

*~2.2 Formation mechanisms and Dynamical planet classes*

Spherical bodies below the hydrogen burning mass limit, about 80 $M_J$, have multiple formation mechanisms and dynamical circumstances. The planetary bodies of the Solar System and most exoplanetary systems formed from materials in a circumstellar proto-planetary disk (see Raymond et al. 2018 for a review). In addition to bodies meeting the IAU definitions for "Planet" and "Dwarf Planet", the population of spherical sub-stellar bodies also includes spherical moons such as the Moon, Ganymede, Callisto, Europa, Io, Titan, Triton, and Charon that orbit planets and dwarf planets. These bodies are "Satellite Planets" – a term that distinguishes these terrestrial bodies from the numerous smaller non-spherical moons with radius less than approximately 200 km. (Stern et al. 2015; Russell 2017). Some ice-rich exoplanets may be scaled up versions of Ganymede (Sotin et al. 2010) or "super-Ganymedes" (Jontof-Hutter 2019). It is possible that Satellite Planets orbiting exoplanets may soon be detectable (Kipping 2021).

Gas giants have three likely formation mechanisms: core accretion, disk instability, and star-like gas collapse. The core accretion and disk instability mechanisms operate in a proto-planetary disk (Boss 2001, 2003; Raymond et al. 2018; Mercer & Stamatellos 2020). Kumar (1963) proposed that the star-like gas fragmentation and collapse of molecular clouds can form bodies below the hydrogen burning mass limit. These bodies are identified as "brown dwarfs".

## 3 Generalized Planetary Composition Classes

The mass-radius relationship suggests the planetary population has three broadly defined composition classes (Chen & Kipping 2017; Otegi et al. 2020): (1) Planets composed of >50% H-He gas by mass with a <50 % by mass fraction of rock and ice, (2) Planets composed of >50% rock or mixed rock-ice by mass with an H-He envelope that is less than 50% by mass, and (3) Planets composed entirely of rock or mixed rock-ice with no H-He envelope. In the Solar System, these three classes are represented by Gas Giants, Ice Giants, and Terrestrial planets respectively. In this section, it is argued that each Solar System planet and all exoplanets may be assigned one of five broad composition classes identified as ***Gas Giant, Rock-Ice Giant***, **gas-rich** ***Terrestrial, Rock Terrestrial, and Rock-Ice Terrestrial*** with the defining characteristics for each class accommodating a wider range of compositions and conditions than found in the Solar System (Table 1). Each of these general composition classes has multiple bulk composition classes that can be characterized with Solar System analog names as will be described in section 5.

*~3.1 Gas Giant Planets and Brown dwarfs*

Gas Giants include planets formed in a proto-planetary disk by core accretion or disk instability mechanisms and brown dwarfs formed by star-like gas collapse. These bodies have an H-He mass percentage greater than 50% by mass (Table 1). The Solar System's Gas Giants, Jupiter and Saturn, most likely formed via core accretion (Raymond et al. 2018). However, the Gas Giant population with masses exceeding 4 $M_{Jup}$ includes planets that may have formed by core accretion, but also formed by disk instability (Santos et al. 2017; Schlaufmann 2018; Adibekyan 2019; Goda & Matsuo 2019) and brown dwarfs formed through star-like gas collapse (Hatzes & Rauer 2015; Caballero 2018; Goda & Matsuo 2019; Luhman & Hapich 2020).

The M-ρ relationship suggests that sub-stellar gas giants over the mass range 0.3 to 60 $M_{Jup}$ follow the same physical relationships whether they formed by core accretion, disk instability, or star-like gas collapse (Hatzes & Rauer 2015). Importantly, it has been noted that there is no breakpoint in the M-ρ relationship at 13 $M_{Jup}$ that would indicate deuterium burning has a significant role in the structure and evolution of gas giant planets and brown dwarfs (Chabrier et al. 2014; Hatzes & Rauer 2015). In addition, star-like gas collapse may form bodies smaller than the deuterium burning limit (e.g. Caballero 2018; Luhman & Hapich 2020). Despite having different formation mechanisms, planets and brown dwarfs from 0.3 to 60 $M_{Jup}$ all appear to share a similar structure and composition (Hatzes & Rauer 2015) and therefore "***Gas Giant***" is an appropriate general composition class for all bodies in this mass range regardless of formation mechanism.

**Table 1: Generalized Planetary Composition Classes**

| Composition class | $f_{H-He}$ (%) | $f_{rock-ice}$ (%) | Mass Range ($M_⊕$) | Radius Range ($R_⊕$) |
|---|---|---|---|---|
| **Gas Giant** | 50.1 – 100.0 | 0.0 – 49.9 | 50 – 19000 | >8.1 |
| **Rock-Ice Giant** | 1.0 – 49.9 | 50.1 – 99.0 | 5 – 106 | 2.3 – 11.3 |
| **Gas-rich Terrestrial** | 0.01 – 0.99 | 99.01 – 99.99 | 4 - 26 | 1.7 – 2.8 |
| **Rock Terrestrial** | <0.01 | 99.99 – 100.00 | <11 | <1.9 |
| **Rock-Ice Terrestrial** | <0.01 | 99.99 – 100.00 | <11 | <2.4 |

*~3.2 Rock-Ice Giant Planets*

Models of the Solar System's Ice Giants, Uranus and Neptune, suggest a possible composition that is ~75-90% rock and ice with the remaining 10-25% of the planetary mass in an outer H-He envelope (dePater & Lissauer 2015; Helled & Fortney 2020). Determination of the mass percentage of water and rock in these planets is difficult due to model degeneracy (Helled & Fortney 2020). The term "ice giant" is derived from models for Uranus and Neptune in which the planetary composition is >50% and possibly as large as 80% ices by mass (dePater & Lissauer 2015, Helled et al. 2020). While there are strong arguments for a significant water ice contribution to the composition of the Solar System's ice giants, the observed parameters of Uranus and Neptune do not exclude models with a rock dominated rather than an ice dominated interior (Teanby et al. 2019; Helled et al. 2020).

Model degeneracy and the large variation in the H-He fraction (e.g. MPB20) of super-Earth to super-Neptune mass exoplanets further complicates the characterization of this class of planets. While some Neptunian planets may have a classic "ice giant" composition, numerous sub-Neptunes can be modeled as rock planets with negligible ices and a radius inflated by an H-He envelope (Lopez & Fortney 2014; Zeng et al. 2019). Planets with this composition can be described as "rock giants" rather than "ice giants" (Teanby et al. 2019; Helled et al. 2020). It has also been shown that the large radii of some "Neptunes" could be explained if Rock-Ice Terrestrial composition planets are inflated with a supercritical steam envelope (Mousis et al. 2020; Aguichine et al. 2021).

The "Neptunian" and "sub-Neptunian" radius planet population is likely to have the following general compositions: (1) a >50% by mass pure rock composition core with a negligible ice mass fraction and the remaining mass in an H-He envelope; (2) a >50% by mass mixed rock-ice core with $f_{rock}$ greater than $f_{ice}$ and the remaining mass in an H-He envelope; (3) a >50% by mass mixed rock-ice core with $f_{ice}$ greater than $f_{rock}$ and the remaining mass in an H-He envelope; and (4) a 100% rock-ice planet with no H-He envelope, but a large radius resulting from a massive supercritical steam envelope that significantly inflates the planetary radius above the pure water composition line (Mousis et al. 2020; Aguichine et al. 2021). This last composition should be grouped with the Terrestrial composition planet classes (section 3.4).

Vazan et al. (2022) found that rock and ice are potentially over 99% miscible in the interiors of planets with masses larger than approximately 1 $M_⊕$. Planets with a bulk composition of mixed rock-ice or rock-ice-gas will not necessarily have distinct rock and ice layers. Instead, these planets could have an interior that is a rock-ice mixture with density gradients and <1% of the planetary mass in the form of an outer water layer or steam atmosphere (Vazan et al. 2022). Given the range of plausible compositions for "Neptunes" and the possibility for high miscibility of rock and ice in the interior of sub-Neptune and Neptune radius planets (Vazan et al. 2022), this class of planets can be comprehensively identified as "***Rock-Ice Giants***" rather than "Ice Giants". The class "***Rock-Ice Giant***" indicates a rock or rock-ice planet with a 1.0 – 49.9% by mass H-He envelope (see Table 1 and section 3.4).

*~3.3 Rock Terrestrial and Rock-Ice Terrestrial Planets*

"Terrestrial" planets have either a 100% rock composition (Rock Terrestrial) or a 100% rock and ice composition (Rock-Ice Terrestrial). The Solar System has four Rock Terrestrial Planets with structure that includes a silicate crust, a silicate mantle, and an iron alloy core (dePater & Lissauer 2015, Zeng et al. 2016). Among the Satellite Planets in the Solar System, the Moon and Io have a Rock Terrestrial composition. The remaining Satellite Planets and most Dwarf Planets have a Rock-Ice Terrestrial composition. Vazan et al. (2022) found that the Solar System's rock-ice

Satellite Planets have insufficient mass for rock and ice to become fully miscible in the interior. These bodies are therefore expected to have an outer icy shell and, depending upon the $H_2O$ and Fe fractions, may have interior liquid ocean, solid ice mantle, rock mantle, and iron core layers (Schubert et al. 2010; de Pater & Lissauer 2015).

### ~3.4 Defining the boundary between Rock-Ice Giant and gas-rich Terrestrial planets

Otegi et al. (2020 – hereafter OBH20) suggested the exoplanet population with mass less than 120 M⊕ may be divided into rock and rock-ice planets versus planets with an H-He envelope using the pure-water composition line in the M-R and M-ρ relationships. There is significant degeneracy in characterizing planets between the pure silicate and pure water composition lines because many possible compositions are consistent with the observed planetary mass and radius. Planets with no H-He envelope falling below the pure-water composition line are Rock or Rock-Ice Terrestrial composition. However, the population of planets with mass and radius values that place them below the pure-water composition line may include planets with an Earth-like pure rock (~30% Fe and ~70% silicate minerals) composition, or mixed rock-ice composition, but with a 0.01 to 4% by mass H-He envelope that inflates the radius (Lopez & Fortney 2014; Zeng et al. 2019; Owen et al. 2020; Armstrong et al. 2020). Some of these may be considered **"gas-rich Terrestrial"** composition planets (Owen et al. 2020; Bean et al. 2021) whereas others are more consistent with a Rock-Ice Giant composition. In this section, we consider how to usefully define the difference between **Rock-Ice Giant** planets and **gas-rich Terrestrial** planets.

The primary advantage of dividing the Terrestrial planet population from Rock-Ice Giant population using the pure-water composition line on the M-R diagram is that any planet above the pure water composition line must have a massive envelope. However, modeling the composition of this envelope suffers from degeneracy as the envelope, may be composed of H-He (Lopez & Fortney 2014; Zeng et al. 2019), H-He with water (Jontoff-Hutter 2019), or a supercritical steam envelope (Mousis et al. 2020; Aguichine et al. 2021). In addition, planets lying below the pure water composition line can still have a significant H-He envelope (Lopez & Fortney 2014; Zeng et al. 2019; Armstrong et al. 2020).

As an alternative to using the pure water composition line to divide the Terrestrial and Rock-Ice Giant populations, an **H-He mass fraction of 1.0% provides a useful composition boundary for dividing Rock-Ice Giant planets from "gas-rich Terrestrial" planets**. Planets with an H-He fraction less than 1.0% are **gas-rich Terrestrial** composition whereas planets with an H-He fraction between 1.0 and 49.9% are **Rock-Ice Giant** composition. Before considering arguments in favor of using a 1.0% H-He mass fraction to divide the Rock-Ice Giant and "gas-rich Terrestrial" planet populations, it is important to note several reasons for expecting some planets with a mass and radius combination below the pure water composition line on a mass-radius diagram will have a low percentage by mass H-He envelope:

~1. Given the nature of planetary building blocks in a proto-planetary disk and planetary formation mechanisms (see review by Raymond et al. 2018), super-Earth mass pure-water composition planets are unlikely to form (Jontof-Hutter 2019). At distances outside the snow line in a proto-planetary disk, pebbles are composed of both rock and ice (Morbidelli et al. 2015). Super-Earth mass and larger planets falling near the pure-water composition line on the M-R diagram are therefore expected to have a rock or mixed rock-ice interior with an outer H-He envelope or supercritical steam envelope that has inflated the planetary radius (Owen et al. 2020; Mousis et al. 2020; Aguichine et al. 2021; Bean et al. 2021). For example, Osborn et al. (2021) note that a gas free model for the planet TOI-755 b ($M_⊕$ = 4.55 +/- 0.62, $R_⊕$ = 2.05 +/- 0.12) requires a water mass fraction of 73 +10/-13 percent that is difficult to explain from a formation point of view. However, models of TOI 755 b with an H-He envelope are consistent with an H-He mass fraction of 0.1 to 1.0 percent on mixed rock-ice or pure rock cores (Osborn et al. 2021).

~2. H-He envelopes may be stripped by stellar irradiation, core powered mass loss, or giant impacts (e.g. Owen et al. 2020; Misener & Schlichting 2021; Gupta & Schlichting 2021; Ogihara et al. 2021) and therefore a planet that forms as a Rock-Ice Giant may evolve toward a rock or rock-ice Terrestrial composition through loss of a significant fraction of the initial H-He envelope (e.g. Dai et al. 2019).

~3. Planets that reach approximately 1 $M_\oplus$ before the gas disk dissipates during planet formation can acquire a 1% by mass H-He envelope (Owen & Mohanty 2016; Owen et al. 2020).

For each of these reasons, the planet population with masses greater than approximately 1.0 $M_\oplus$ should include some gas-rich Terrestrial planets with a low percentage by mass H-He envelope. However, there are several drawbacks to using the pure water composition line *to define* the boundary between these gas-rich Terrestrial planets and Rock-Ice Giant planets:

~Since pure-water composition planets are generally not expected to form (Jontof-Hutter 2019; Bean et al. 2020), the pure water composition line on the M-R diagram is unlikely to represent the actual composition of observed super-Earth to Neptune mass exoplanets.

~The H-He mass fraction required to account for a pure-water composition radius increases from less than 0.1% for small Terrestrial planets to over 3% at 20 to 25 Earth masses depending upon the composition of the core (Lopez & Fortney 2014; Zeng et al. 2019). As an extreme example, TOI-849 b, which has a mass of 39.09 +2.66/-2.55 $M_\oplus$ and radius of 3.44 +0.16/-0.12 $R_\oplus$ (Armstrong et al. 2020), lies below the pure-water composition line. Using the pure water composition criterion, TOI-849 b would be classified as a Terrestrial planet. However, TOI-849 b is likely the remnant core of a super-Neptune or Gas Giant that retains a 2.8 - 3.9% by mass H-He envelope (Armstrong et al. 2020). While TOI-849 b falls below the pure-water composition line on the M-R diagram, this planet does not have a terrestrial composition and instead matches the Rock-Ice Giant composition class.

~Planets with an H-He envelope are subject to radius inflation resulting from a young age (Lopez & Fortney 2014; Libby-Roberts et al. 2020), increasing stellar flux or equilibrium temperature (Lopez & Fortney 2014; Zeng et al. 2019), and as a result of tidal inflation (MPB20). Mechanisms that inflate the radius of a planet can result in identical composition planets falling on opposite sides of the pure water composition line. For example, a 1 Gyr age, 8.5 $M_\oplus$ planet with a 1% by mass H-He envelope will have radii of 2.31 $R_\oplus$ or 2.58 $R_\oplus$ for stellar flux values of 10 $F_\oplus$ and 1000 $F_\oplus$ respectively (Lopez & Fortney 2014). The pure water composition radius for an 8.5 $M_\oplus$ planet is 2.49 $R_\oplus$ (Zeng et al. 2016) and therefore this planet would be assigned the Terrestrial composition class for a stellar flux of 10 $F_\oplus$, or the Rock-Ice Giant composition class for a stellar flux of 1000 $F_\oplus$.

Using models of Zeng et al. (2019), increasing the temperature of a 4.0 $M_\oplus$ Earth-like composition planet with a 1.0% $H_2$ envelope from 300K to 500K would shift the composition classification from gas-rich Terrestrial to Rock-Ice Giant if the classification is based upon the radii of the pure water composition line. Another possible mechanism for this type of classification inconsistency is tidal inflation of an H-He envelope which can significantly inflate the radii of planets with orbital periods less than 12 days (see analysis and Table 3 of MPB20).

~Rock-ice Terrestrial composition planets with no H-He envelope can have a steam envelope that inflates the planetary radius above the pure water line if the water mass fraction of the planet reaches approximately 25 - 50% (Aguichine et al. 2021).

As an alternative to the pure water composition line, it is here proposed that planets with 0.01 – 0.99% by mass H-He gas be classified as **gas-rich Terrestrial** planets and planets with 1.0 – 49.9% by mass H-He gas be classified as **Rock-Ice Giants**. In contrast to the pure water composition line, defining the boundary between gas-rich Terrestrial and Rock-Ice Giant planets using a 1.0% H-He mass fraction **provides a consistent composition boundary in the classification of planets that is not altered by inflation mechanisms**.

Given model degeneracy and uncertainty in mass and radius measurements, determining the H-He fraction for individual planets close to the pure water composition line may prove difficult in some cases. Nonetheless, the difference between a Gas Giant and a Rock-Ice Giant is defined based upon the H-He mass fraction and it therefore brings consistency to also define the difference between gas-rich Terrestrial and Rock-Ice Giant planets using a specified H-He mass fraction (Table 1).

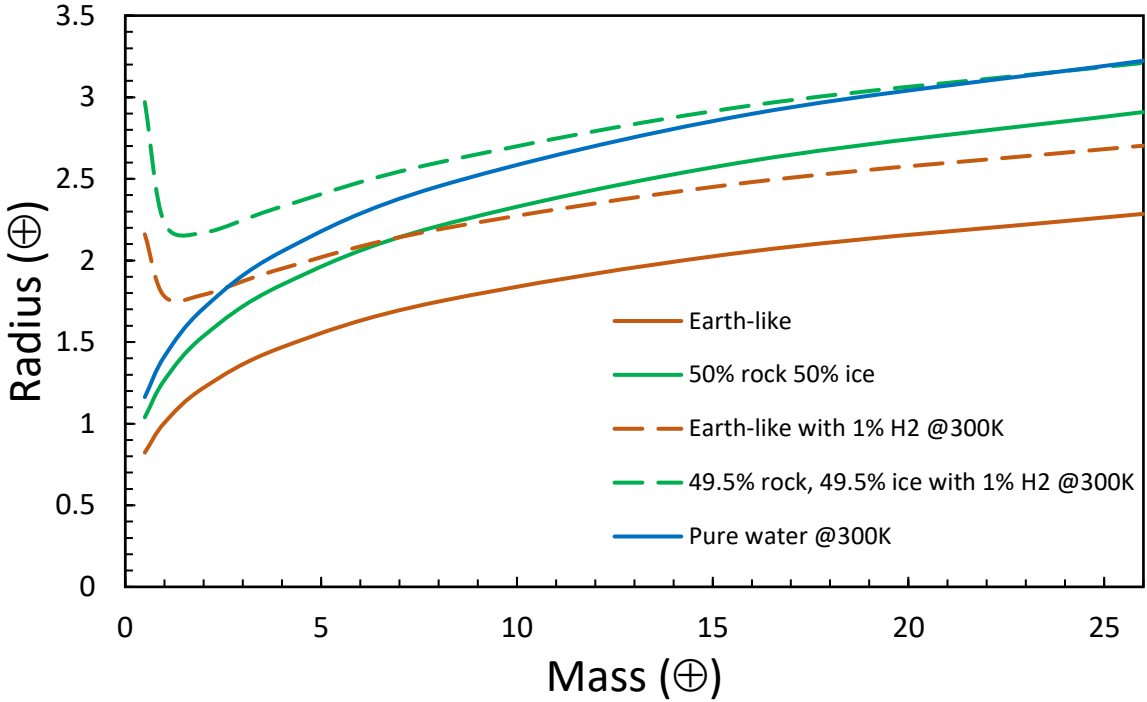

**Figure 1:** Mass-Radius-composition relationships for planets with mass < 26 $M_\oplus$ using Zeng et al. 2019 models. Composition curves shown are Earth-like (solid brown), Earth-like with 1% $H_2$ at 300K (dashed brown), 50% rock – 50% water (solid green), 49.5% rock and 49.5% water with 1% $H_2$ at 300K (dashed green), and pure water at 300K. (solid blue). In the range 7 to 26$M_\oplus$ planets with 50% rock – 50% water to 100% water composition fall between the 300K 1% $H_2$ models.

The selection of 1.0% by mass as the boundary between gas-rich Terrestrial and Rock-Ice Giants is supported by the comparison of several mass-radius-composition curves in Figure 1. The solid lines in Figure 1 provide Earth-like, 50% rock – 50% water, and 100% water composition M-R curves for 0.5 to 26.0 $M_\oplus$ planets (Zeng et al. 2016, 2019). Also plotted are 300K models for an Earth-like core with a 1.0% by mass $H_2$ envelope and a 49.5% rock – 49.5% water core with a 1.0% by mass $H_2$ envelope (Zeng et al. 2019). It is evident in Figure 1, that the 300K 1.0% $H_2$ envelope models reasonably bracket the envelope free 50% rock – 50% $H_2O$ to 100% $H_2O$ composition range. Planets that fall in this range on the M-R diagram are more likely to contain a <1% by mass H-He envelope with a higher rock fraction than to have the envelope free high water mass fraction composition matched by their location on the M-R diagram (e.g. Osborn et al. 2021). Note that the selection of 300K models provides radii close to the minimum radii for planets with 1% by mass H-He as increasing stellar irradiation (Lopez & Fortney 2014; Zeng et al. 2019) and tidal inflation effects (MPB20) will both increase the observed planetary radii. The 300K models therefore provide radii similar to what might be expected for planets with 1% H-He by mass at greater orbital distances, or potential "cool gas-rich Terrestrial" planets.

Aguichine et al. (2021) developed M-R relationships for irradiated ocean planets using the water EoS model of Mazevet et al. (2019). These models include a refractory interior, condensed fluid $H_2O$ layer, a steam atmosphere and no H-He envelope. Following the Aguichine et al. (2021) models, planets with a steam envelope and water mass fractions of 20% and 50% have similar radii to planets that are 50% and 100% liquid water by mass respectively in the 10-20 $M_\oplus$ range.

*~3.5 Summary*

Every Solar System Planet, Dwarf Planet, and Satellite Planet, and all exoplanets can be assigned one of five generalized composition classes here identified as ***Gas Giant***, ***Rock-Ice Giant***, ***gas-rich Terrestrial***, ***Rock Terrestrial***, or ***Rock-Ice Terrestrial***. The composition characteristics and empirical mass and radius ranges for each class are summarized in Table 1.

Each of these general composition classes has multiple possible bulk composition classes that will be described and characterized with Solar System analog names in section 5. It is important to emphasize that the composition class names do not imply that the rock, ice, and gas components must exist in specific states. Interior models are constantly changing with new analyses and equation of state models (e.g. Helled & Stevenson 2017; Hakim et al. 2018b, 2019a; Mazevet et al. 2019; Vazan et al. 2022). For example, interior models for Gas Giant and Rock-Ice Giant planets now incorporate a "fuzzy core" (e.g. Helled & Stevenson 2017; Helled et al. 2020). Recent EoS models for silicon carbide indicate that SiC is unstable in the interiors of carbon-enriched super-Earth mass planets and therefore carbon enriched exoplanets have silicate, rather than the SiC mantles found in earlier models for carbon enriched planets (Hakim et al. 2018b, 2019a). Since the broad composition classes are dependent only upon the mass fractions of H-He gas, Fe and silicate rock, and astrophysical ices, the classes will not require constant modification as new EoS and interior structure models are developed.

**Section 4: Observed mass limits and overlap ranges for Terrestrial, Rock-Ice Giant, and Gas Giant composition classes**

In this section, mass-radius data is examined to determine the observed upper and lower mass limits for the Rock and Rock-Ice Terrestrial, gas-rich Terrestrial, Rock-Ice Giant, and Gas Giant composition classes. The upper and lower limits also help identify the overlapping mass ranges for Terrestrial/Rock-Ice Giant and Rock-Ice Giant/Gas Giant composition classes. In order to examine this question the following samples and models are used:

~1. Exoplanet sample from OBH20 and references therein which includes exoplanets with mass <120 $M_\oplus$, mass uncertainty <25%, and radius uncertainty <8%.

~2. Confirmed exoplanets from the NASA exoplanet archive (Akeson et al. 2013) meeting the OBH20 standards for mass and radius uncertainty.

~3. Exoplanet sample and analysis from MPB20 which includes sub-Saturn radius (4.0 to 7.5 $R_\oplus$) planets. MPB20 provided improved H-He envelope mass fractions for the planets in their sample by accounting for tidal inflation. For planets with a gas envelope, MPB20 find that the actual envelope mass fraction at a given radius is lower in their tidal model than the envelope mass fractions determined from non-tidal models such as Lopez & Fortney (2014).

~4. Mass-Radius relations for Terrestrial planets with compositions ranging from pure Fe to pure $H_2O$ (Zeng et al. 2016, 2019).

~5. Mass-Radius and gas envelope fraction relations developed by Lopez & Fortney (2014) and Zeng et al. (2019).

~6. Mass-Radius H-He envelope fraction relations for Gas Giants from Fortney et al. (2007) and Baraffe et al. (2008).

The exoplanet sample from OBH20 covers the mass range necessary to explore the upper mass limit for Terrestrial planets, the lower and upper mass limits for Rock-Ice Giants, and the lower mass limit for Gas Giants. It is important to keep in mind mass limits identified from this sample are empirical limits based upon currently known exoplanet samples with highly accurate mass and radius measurements and therefore could be adjusted with future exoplanet discoveries and reductions in mass and radius measurement uncertainty.

The upper mass limit for Gas Giants and the lower mass limit for Terrestrial planets are not covered by the OBH20 sample. For Brown dwarfs formed by star-like gas collapse mechanisms, the upper mass limit is approximately 60 $M_{Jup}$ (Hatzes & Rauer 2015) whereas the upper mass limit for planets formed by core accretion or disk instability in a proto-planetary disk is approximately 10-25 $M_{Jup}$ (Goda & Matsuo 2019).

The lowest mass Terrestrial planet in the OBH20 sample is TRAPPIST-1 d (0.297 $M_\oplus$). The lower mass limit for attaining a spheroidal shape, based upon the population of planetary bodies in the Solar System, is approximately 3.75 x $10^{19}$ kg, the mass of Saturn's Satellite Planet Mimas (Lineweaver & Norman 2010; Tancredi 2010; Russell 2017).

*~4.1 Upper Mass Limit for Terrestrial Planets*

The planets identified for the sample of OBH20 and NASA Exoplanet archive may be compared to M-R composition relationships provided by Zeng et al. (2016, 2019). The most massive planet in the OBH20 sample falling below the pure water composition line is Kepler-411 b ($M_\oplus$ =25.76+/- 2.544, $R_\oplus$ = 2.40 +/- 0.056) and sets the observed upper mass limit for the Rock Terrestrial planet class at 26 $M_\oplus$. Using the tables for the Zeng et al. (2016, 2019) M-R relations, Kepler-411 b could have an envelope free composition that is approximately 10% Fe and 90% silicates. Given the relatively low iron fraction of this planet if it is a Rock-Terrestrial planet, Kepler-411b could instead be an Earth-like composition core gas-rich Terrestrial planet with 0.01-0.1% by mass H-He envelope (Lopez & Fortney 2014; Zeng et al. 2019).

Kepler-131b ($M_\oplus$ = 16.13 +/- 3.5, $R_\oplus$ = 2.1 +0.2/-0.1) is another massive Rock Terrestrial or gas-rich Terrestrial planet with an H-He envelope mass fraction less than 1.0%. From Zeng et al. (2016) Kepler-131b has mass and radius that match a 20% Fe and 80% silicate minerals Rock Terrestrial planet with no H-He envelope. Following Lopez & Fortney (2014), Kepler-131b could also be a gas-rich Terrestrial planet with 30% Fe and an approximately 0.1% H-He envelope.

The five largest mass Terrestrial planets in the OBH20 sample and NASA Exoplanet Archive with radii too small to accommodate an Earth-like core gas-rich Terrestrial planet composition solution, and therefore are likely to be gas envelope free are TOI 1634 b (10.14 +/-0.95 $M_\oplus$ - Hirano et al. 2021), Kepler 20b (9.7 +/- 1.4 $M_\oplus$), Kepler 107c (9.39 +/- 1.77 $M_\oplus$), HD 213885b (8.83 +0.66/-0.65 $M_\oplus$), and K2-216b (8.0 +/- 1.6 $M_\oplus$). This suggests a possible upper mass limit for a pure Fe-silicate rock composition planet of ~10 $M_\oplus$. Terrestrial planets in the 10 – 26 $M_\oplus$ mass range are more likely to be gas-rich Terrestrial planets possessing a <1% by mass H-He envelope. In the OBH20 sample, there are only seven Terrestrial composition planets with masses exceeding 10 $M_\oplus$. Only Kepler-411 b and Kepler-131 b have radii smaller than expected for a pure-silicate rock composition supporting the possibility that most, if not all, Terrestrial planets exceeding 10 $M_\oplus$ are gas-rich Terrestrial composition planets.

*~4.2: A sample of Candidate "gas-rich Terrestrial" Composition Planets*

The NASA Exoplanet Archive was searched for planets meeting the OBH20 standard for mass uncertainty (<25%) and radius uncertainty (<8%) to identify candidate gas-rich Terrestrial composition planets. Composition classification was determined using the mass-radius-composition models of Zeng et al (2019).

Data necessary for this classification analysis includes planetary mass, radius, and stellar flux or equilibrium temperature. The relevant models from Zeng et al. (2019) include planets with 0.1%, 0.3%, and 1% $H_2$ envelopes on an Earth-like core or on a 49.5% rock - 49.5% $H_2O$ core at equilibrium temperatures of 300K, 500K, 700K, and 1000K. For a given set of mass-radius values, a gas-rich Terrestrial planet with an Earth-like core will have a higher $H_2$ mass fraction than a planet with a 49.5% rock – 49.5% $H_2O$ core. Based upon the Zeng et al. (2019) models, all planets in Table 2 have < 1% $H_2$ by mass on both rock cores and rock-ice cores. Planets that have mass-radius values consistent with <1% $H_2$ on a rock-ice core, but >1% $H_2$ on an Earth-like composition core are not included in Table 2 as any such planets are possible Rock-Ice Giants.

*Table 2: Candidate gas-rich Terrestrial Planet Sample*

| Planet | Mass (M$_\oplus$) | Radius (R$_\oplus$) | S$_\oplus$ / Teq (K)[a] | Mass –Radius Reference |
|---|---|---|---|---|
| **TOI-776 b** | 4.0 +/-0.9 | 1.85 +/-0.13 | 11.5 / 514 | Luque et al. 2020 |
| **GJ 9827 d** | 4.04 +0.82/-0.84 | 2.022 +.046/-.043 | - / 646 | Rice et al. 2019 |
| **Kepler 60 d** | 4.16 +0.84/-0.75 | 1.99 +/-0.16 | 161 / - | Jontof-Hutter et al. 2016 |
| **Kepler 60 b** | 4.19 +0.56/-0.52 | 1.71 +/-0.13 | 318 / - | Jontof-Hutter et al. 2016 |
| **TOI-755 b** | 4.55 +/-0.62 | 2.05 +/-0.12 | 731 / 1371 | Osborn et al. 2021 |
| **TOI-270 d** | 4.78 +/-0.43 | 2.133 +/-0.058 | - / 387 | VanEylen et al. 2021 |
| **HD 39091 c** | 4.82 +0.86/-0.84 | 2.05 +/-0.50 | 309 / 1170 | Huang et al. 2018 |
| **K2-111 b** | 5.29 +0.76/-0.77 | 1.82 +0.11/-0.09 | - / 1309 | Mortier et al. 2020 |
| **K2-146 b** | 5.77 +/-0.18 | 2.05 +/-0.06 | 20.7 / 534 | Hamann et al. 2019 |
| **TOI-1201 b** | 6.28 +0.84/-0.88 | 2.415 +0.091/-0.090 | 40.6 / 703 | Kossakowski et al. 2021 |
| **K2-138 c** | 6.31 +1.13/-1.23 | 2.299 +0.120/-0.087 | 249 / 1012 | Lopez et al. 2019 |
| **Kepler 454 b** | 6.84 +/-1.40 | 2.37 +/- 0.13 | 116 / - | Gettel et al. 2016 |
| **TOI-421 b** | 7.17 +/-0.66 | 2.68 +0.19/-0.18 | 155 / 981 | Carleo et al. 2020 |
| **Kepler 307 b** | 7.44 +0.91/-0.87 | 2.43 +/- 0.09 | 59.7 / - | Jontof-Hutter et al. 2016 |
| **TOI-178 f** | 7.72 +1.67/-1.52 | 2.287 +0.108/-0.110 | - / 521 | Leleu et al. 2021 |
| **HD 5278 b** | 7.8 +1.5/-1.4 | 2.45 +/- 0.05 | 132 / 943 | Sozzetti et al. 2021 |
| **K2-138 d** | 7.92 +1.39/-1.35 | 2.39 +0.104/-0.084 | 147 / 888 | Lopez et al. 2019 |
| **HD 97658 b** | 8.3 +/-1.1 | 2.12 +/-0.06 | - / 751 | Ellis et al. 2021 |
| **TOI-1260 b** | 8.60 +/-1.50 | 2.34 +/-0.11 | 91 / 860 | Georgieva et al. 2021 |
| **Kepler 102 e** | 8.93 +/-2.0 | 2.22 +/-0.07 | 20 / - | Marcy et al. 2014 |
| **K2-285 b** | 9.68 +1.21/-1.37 | 2.59 +/-0.06 | 234 / 1089 | Palle et al. 2019 |
| **TOI-763 b** | 9.79 +/-0.78 | 2.28 +/-0.11 | - / 1038 | Fridlund et al. 2020 |
| **TOI-1062 b** | 10.15 +0.81/-0.84 | 2.265 +0.96/-0.91 | - / 1077 | Otegi et al. 2021 |
| **Kepler 538 b** | 10.6 +2.5/-2.4 | 2.215 +0.040/-0.034 | 2.99 / 380 | Mayo et al. 2019 |
| **K2-263 b** | 14.8 +/-3.1 | 2.41 +/-0.12 | - / 470 | Mortier et al. 2018 |
| **Kepler 131 b** | 16.13 +/-3.5 | 2.41 +/-0.20 | 66 / - | Marcy et al. 2014 |
| **HIP 97166 b** | 20.0 +/-1.5 | 2.74 +/-0.13 | - / 757 | MacDougall et al. 2021 |
| **GJ 143 b** | 22.7 +2.2/-1.9 | 2.61 +0.17/-0.16 | - / 422 | Drogomir et al. 2019 |
| **K2-292 b** | 24.5 +/-4.4 | 2.63 +0.11/-0.10 | 67 / 795 | Luque et al. 2019 |
| **Kepler 411 b** | 25.6 +/-2.6 | 2.401 +/-0.053 | - / 1138 | Sun et al. 2019 |

Note a – stellar flux and equilibrium temperatures identified from multiple sources in the NASA Exoplanet archive.

The population of candidate gas-rich Terrestrial composition planets identified from the current exoplanet sample, meeting the OBH20 standards for mass and radius uncertainty, and with stellar flux or equilibrium temperature values available in the NASA Exoplanet Archive, has a mass range of 4.0 – 25.6 M$_\oplus$ and a radius range of 1.70 to 2.75 R$_\oplus$. Many of the planets listed in Table 2 fall into a radius range often identified as "mini-Neptunes". However, based upon the composition analysis, these planets are instead examples of "gas-rich super-Earth" composition planets. This class of planet which has a voluminous, but low percentage by mass H-He envelope on a Terrestrial core, represents the first fundamentally new type of planet that has been identified in exoplanet studies (Bean et al. 2021).

*~4.3 Lower Mass Limit for Rock-Ice Giants*

As defined in this paper, the H-He envelope mass fraction of a Rock-Ice Giant planet is between 1.0 and 49.9 % by mass. The lowest mass Rock-Ice Giant planets from the OBH20 sample, the MPB20 sample, and the list of confirmed planets from the NASA Exoplanet archive which meet the OBH20 standard for mass and radius uncertainty are listed in Table 3. Most of these planets have masses between 6.2 M$_\oplus$ and 6.9 M$_\oplus$. The lowest mass planet that must have an H-He envelope fraction exceeding 1.0% using the Zeng et al. (2016, 2019) composition models is Kepler 26 b with a mass of 5.12 +0.65/-0.61 M$_\oplus$ (Jontof-Hutter et al. 2016). The observed mass range for the lower mass limit of Rock-Ice Giant composition planets is therefore 5-7 M$_\oplus$.

**Table 3: Lowest mass Rock-Ice Giants**

| Planet | Mass (M⊕) | Radius (R⊕) | Reference |
|---|---|---|---|
| **Kepler 26b** | 5.12 +0.65/-0.61 | 2.78 +/-0.11 | Jontof-Hutter et al. 2016 |
| **TOI-561 c** | 5.40 +/-0.98 | 2.878 +/- 0.096 | Lacadelli et al. 2020 |
| **Kepler 177 b** | 5.84 +0.86/-0.82 | 3.50 +0.19/-0.15 | Vissapragada et al. 2020 |
| **Kepler 26 c** | 6.20 +/-0.65 | 2.72 +/-0.12 | Jontof-Hutter et al. 2016 |
| **LTT 3780 c** | 6.29 +0.63/-0.61 | 2.42 +/- 0.10 | Nowak et al. 2020 |
| **Kepler 79e** | 6.3 +/-1.0 | 3.414 +/- 0.129 | Yoffe et al. 2020 |
| **Kepler-87 c** | 6.4 +/- 0.8 | 6.14 +/- 0.29 | Millholland et al. 2020 |
| **TOI-125 c** | 6.63 +/-0.99 | 2.759 +/-0.10 | Nielsen et al. 2020 |
| **Kepler-11 e** | 6.7 +1.2/-1.0 | 4.0 +0.2/-0.3 | Hadden & Lithwick 2017 |
| **Kepler 80 c** | 6.74 +1.23/-0.86 | 2.74 +/-0.12 | MacDonald et al. 2016 |
| **Kepler-11 d** | 6.8 +0.7/-0.8 | 3.3 +/-0.2 | Hadden & Lithwick 2017 |
| **Kepler 80 b** | 6.93 +1.05/-0.70 | 2.67 +/- 0.10 | MacDonald et al. 2016 |
| **Kepler-36 c** | 7.13 +/-0.18 | 3.679 +/-0.098 | Vissapragada et al. 2020 |
| **HD 191939 c** | 7.2 +/-1.4 | 3.08 +/- 0.07 | Lubin et al. 2022 |
| **Kepler-223 d** | 8.0 +1.5/-1.3 | 5.24 +0.26/-0.45 | Mills et al. 2016 |
| **GJ 1214 b** | 8.17 +/- 0.43 | 2.742 +/-0.05 | Cloutier et al. 2021 |

*~4.4 Upper Mass Limit for Rock-Ice Giants and Lower mass limit for Gas Giants*

The tidal inflation model of MPB20 complicates determination of the upper mass limit for Rock-Ice Giants and lower mass limit for Gas Giants. The results of MPB20 demonstrate that planets with large radii consistent with an uninflated Gas Giant composition, falling at or above the 50% H-He line on the M-R diagram, can have a tidally inflated radius. Many of these tidally inflated planets will therefore actually be Rock-Ice Giant composition with an H-He mass fraction less than 50 percent. Tidal inflation has a greater impact on planets with shorter orbital periods (MPB20). Planets in the MPB20 sample with orbital periods < 12 days have H-He envelope fractions after accounting for tidal inflation that are only 25-50% of the value derived from a non-tidal model (see Table 3 of MPB20). Most planets in the OBH20 sample from 40 - 120 M⊕ have orbital periods <6 days and therefore should experience H-He envelope tidal inflation.

The tidal model of MPB20 did not explore planets with actual fenv > 50%. Based upon the models of Fortney et al. (2007) and Baraffe et al. (2008), planets with 50% H-He and 90% H-He and with mass ranging from 50 – 100 M⊕ should have radii of 8.1-8.3 R⊕ and 11.4 R⊕ respectively for a 5 Gyr and 0.045 AU model. If the results of MPB20 extrapolate to radii larger than sub-Saturn radius planets, many short orbital period planets with non-tidally inflated model gas fractions greater than 50% are likely to have actual gas mass fractions less than 50%. Any such planets would be inflated Rock-Ice Giants rather than Gas Giants. As an example, MPB20 demonstrated that WASP-107 b (R⊕ = 10.6 +/- 0.3), which has a non-Tidal envelope fraction of approximately 75%, could have an actual envelope fraction as low as 10% after accounting for tidal inflation.

For the OBH20 sample with orbital periods <12 days, planets with radii < 9.9 R⊕, or about 60% H-He with models that do not account for tidal inflation (Fortney et al. 2007; Baraffe et al. 2008), will be considered Rock-Ice Giants. Planets with radii from 9.9 – 11.4 R⊕ will be considered possible Rock-Ice Giants, and planets with radii >11.4 R⊕ will be considered Gas Giants.

The most massive planet from the OBH20 sample with radius less than 8.1 R⊕, and therefore must be a Rock-Ice Giant regardless of tidal inflation effects is CoRoT-8 b (M⊕ = 69.3 +/- 10.8; R⊕ = 6.94 +0.18 -0.19 – Raetz et al. 2019). The most massive short orbital period planets from the OBH20 sample with radii less than 9.9 R⊕ have masses from 100 - 106 M⊕ (K2-287 b, HD 149026 b, and K2-295 b). The upper mass limit for Rock-Ice Giants is therefore at least 70 M⊕ and likely 100-110 M⊕.

The least massive Gas Giants in the OBH20 sample with radii exceeding 11.4 $R_⊕$ have masses from 52 - 57 $M_⊕$ (HAT-P-48 b, KELT-11 b, and WASP-127 b). Therefore, the mass overlap range for Rock-Ice Giant and Gas Giant planet populations is approximately 50 – 110 $M_⊕$.

**Section 5: Solar System Analog Names for Characterizing Planetary Bulk Composition Classes**

The classes *Rock Terrestrial*, *Rock-Ice Terrestrial*, *gas-rich Terrestrial*, *Rock-Ice Giant*, and *Gas Giant* provide a broad characterization of the rock, ice, and gas composition of planets (Table 1). However, within each class there can be important differences in composition. For example, Rock Terrestrial planets can be "Earth-like" with a silicate fraction > 50% or "Mercury-like" with iron fraction >50%. Rock-Ice Giants and Rock-Ice Terrestrial planets will have varying fractions of rock and ice ranging from mostly rock to mostly ice, while Gas Giants can have Z>2 element fractions ranging from <10% to nearly 50%. Table 4 provides a list of planetary bulk compositions and a system Solar System analog names for characterizing the bulk composition classes. Table 5 provides examples of the Solar System analog names applied to exoplanets.

**Table 4: Planet Composition Classes with Solar System Analog Names**

| Composition Class | $f_{H-He}$ (%) | $f_{R-I}$ (%) | Mass Range ($M_⊕$) |
|---|---|---|---|
| ***Gas Giants*** | 50.1 - 100.0 | 0.0 - 49.9 | 50 - 19000 |
| **Brown Dwarf** | 90.0 - 100.0 | <10.0 | 4100 - 19000 |
| **Super-Jupiter** | 90.0 – 100.0 | <10.0 | 1300 - 4100 |
| **Jupiter** | 80.1 - 100.0 | 0.0 – 19.9 | 50 - 1300 |
| **Saturn** | 50.1 – 80.0 | 20.0 – 49.9 | 50 - 300 |
|  |  |  |  |
| ***Rock-Ice Giants*** | 1.0 – 49.9 | 50.1 – 99.0 | 5 - 110 |
| **Super-Neptune** |  |  | 50 - 110 |
| **Neptune** |  |  | 10 – 50 |
| **Sub-Neptune** |  |  | 5 - 10 |
|  |  |  |  |
| ***Gas-rich Terrestrial*** | 0.01 – 0.99 | 99.01 – 99.99 | 4 - 26 |
| **Gas-rich super-Mercury** |  |  |  |
| **Gas-rich super-Earth** |  |  |  |
| **Gas-rich super-Europa** |  |  |  |
| **Gas-rich super-Ganymede** |  |  |  |
|  |  |  |  |
| ***Rock-Terrestrial*** | <0.01 | >99.99 | <11 |
| **Earth  a** |  |  | <4 |
| **Mercury  b** |  |  | <4 |
| **Carbon-rich Earth  c** |  |  | <4 |
| **Super-Earth** |  |  | 4 – 11 |
| **Super-Mercury** |  |  | 4 – 11 |
| **Carbon-rich super-Earth** |  |  | 4 – 11 |
|  |  |  |  |
| ***Rock-Ice Terrestrial*** | <0.01 | >99.99 | <11 |
| **Ganymede  d** |  |  | <4 |
| **Super-Ganymede** |  |  | 4 – 11 |
| **Europa  e** |  |  | <4 |
| **Super-Europa** |  |  | 4 – 11 |

Notes: a – Earth class planets have >50% silicates and <50% iron by mass, b – Mercury class planets have > 50% iron and < 50% silicates by mass, c – carbon-rich Earth planets have a carbon saturated rock mineralogy, d – Ganymede class planets are >10% ice and <90% rock by mass, e – Europa class planets have 0.2 – 10% ice and 90.0-99.8% rock by mass. Rock Terrestrial planets are <0.01% H-He and <0.2% ice by mass.

*~5.1 Rock, Rock-Ice, and gas-rich Terrestrial Composition Classes*

Rock Terrestrial planets can be divided into three bulk composition classes which indicate rock planets that are >50% iron (Mercuries), >50% silicates (Earths), and carbon enriched (carbon-rich Earths). Earth, Mercury, and carbon-rich Earth planets are those with mass < 4 Earth masses. Rock Terrestrial planets with masses from 4 – 11 $M_\oplus$ overlap the mass range for gas-rich Terrestrial planets (Section 4.2) and are characterized as super-Mercury, super-Earth, or carbon-rich super-Earth composition planets. Adibekyan (2021) has identified a list of super-Mercury class planets with >60% Fe by mass.

Carbon-rich Earth/super-Earth planets are carbon-saturated and more likely to form around stars with C/O ratios > 0.8 (Bond et al. 2010). Fortney (2012) found that 10-15% of stars may have a C/O ratio > 0.8. However, more recent studies have found 100% of stars in the samples with C/O ratios < 0.8 (Suárez-Andrés et al. 2018; Bedell 2018) indicating carbon-enriched planets could be rare. Bond et al. (2010) presented models for carbon-enriched planets as "carbide planets" with an iron-carbide alloy core, and a mantle composed of SiC and TiC overlain by a graphite crust. Allen-Sutter et al. (2020) found that delivery of water to a carbide planet could result in a silicate crust and diamond with hydrous silica overlying a mantle of SiC. However, from high temperature and pressure experiments, Hakim et al. (2018b, 2019a) found that SiC is unstable in the interiors of larger rocky exoplanets unless the conditions are extremely reducing. Hakim et al. (2019a) presented models for the interior of carbon-enriched rocky exoplanets that include a metallic iron alloy core, a silicate (pyroxene and olivine) mantle with no SiC, and a graphite crust. Once a carbon-enriched planet reaches C-saturation, additional carbon does not alter the silicate mineralogy but simply adds to the graphite crust with an increasing possibility of a diamond-silicate upper mantle (Hakim et al. 2019a).

Rock-Ice Terrestrial planets can be divided into Ganymede-like and Europa-like bulk composition classes. Ganymede composition class planets have >10% ices by mass whereas Europa composition class planets have 0.2 – 10% ices by mass. As with Rock-Terrestrial planets, super-Ganymede and super-Europa class planets overlap the gas-rich Terrestrial mass range from 4 – 11 $M_\oplus$, whereas Ganymede and Europa class planets are < 4 $M_\oplus$.

Lingam & Loeb (2019) found that in order to cover all land masses on Earth a minimum of 4.3 oceans of surface water is required – or approximately 6.3 total oceans after accounting for the Earth's estimated mass of mantle water. Adopting 1.4 x $10^{21}$ kg per ocean indicates that an Earth-mass planet with approximately 0.15% by mass water could be an "ocean world" free of exposed land mass. Therefore, a minimum planetary water mass fraction of 0.2% is adopted for Rock-Ice Terrestrial composition planets. The bulk composition of Europa is over 90% silicates and iron but includes an ice crust and interior ocean layer with a total thickness slightly over 100 km (Vance et al. 2017). Europa class planets therefore will have 0.2 – 10% of the total planetary mass as water and other ices. Ganymede and super-Ganymede class planets have >10% ices by mass.

Gas-rich Terrestrial planets have masses ranging from 4.0 – 26.0 $M_\oplus$ (section 4.2). These planets are rock or rock-ice Terrestrial composition planets with a 0.01 – 0.99 percent by mass H-He envelope. The bulk composition classes are "gas-rich super-Earth", "gas-rich super-Mercury", "gas-rich super-Europa", and "gas-rich super-Ganymede". Given model degeneracy, additional data beyond mass, radius, and equilibrium temperature (Teq) will be needed to distinguish between the possible gas-rich Terrestrial bulk composition classes.

*~5.2 Rock-Ice Giant composition classes*

Rock-Ice Giants have H-He envelopes that account for 1.0 – 49.9 percent of the planetary mass. The cores of Rock-Ice Giants can have varying rock-ice fractions including: (1) a pure rock composition; (2) a mixed rock-ice composition with $f_{rock}$ greater than $f_{ice}$; and (3) a mixed rock-ice composition with $f_{ice}$ greater than $f_{rock}$. The traditional "Ice Giant" composition models for Uranus and Neptune (dePater & Lissauer 2015) would be consistent with the third composition type. However, a "Rock Giant" composition, consistent with the first two composition types cannot be ruled out for Uranus and Neptune (Helled & Fortney 2020; Teanby et al. 2020). As an example from the exoplanet population, the planets of the TOI-125 system, TOI-125b, TOI-125c and TOI-125d, are modeled by Nielsen et al. (2020) to have 2.0 - 4.1% H-He gas, 62-70% rock, and 32-36% water which is consistent with "Rock Giant" composition.

**Table 5: Composition Classification Applied to Exoplanets**

| Planet | Period (d) | Mass (M⊕) | Radius (R⊕) | Teq (K) | Reference | Classification |
|---|---|---|---|---|---|---|
| **LHS 1140c** | 3.77 | 1.77 +.17/-.16 | 1.169 +/-0.037 | 709 | Lillo-Box et al. 2020a | Earth |
| **GJ 486b** | 1.47 | 2.82 +0.11/-.12 | 1.305 +/-0.06 | 701 | Trifanov et al. 2021 | Earth |
| **Kepler 36 b** | 13.84 | 5.56 +0.41/-0.45 | 1.582 +/-0.015 | 978 | Yofee et al. 2021 | Super-Earth |
| **K2-291 b** | 2.225 | 6.4 +/-1.1 | 1.582 +/-0.04 | | Dai et al. 2019 | Super-Earth |
| **HD213885 b** | 1.008 | 8.83 +/-0.66 | 1.745 +/-0.052 | 2128 | Espinoza et al. 2020 | Super-Earth |
| **K2-229 b** | 0.58 | 2.49 +.43/-.45 | 1.197 +/-0.04 | 1818 | Dai et al. 2019 | Mercury |
| **HD 137496 b** | 1.62 | 4.04 +/-0.55 | 1.31 +.06/.05 | 2130 | Azevedo Silva et al. 2022 | Super-Merury |
| **Kepler 107 c** | 4.901 | 9.39 +/-1.77 | 1.597+/-0.026 | 1379 | Bonoma et al. 2019 | Super-Mercury |
| **L98-59 c** | 3.69 | 2.22 +0.26/-0.25 | 1.385 +.095/-.075 | 553 | Demangeon et al. 2021 | Europa |
| **HD136352 b** | 11.58 | 4.72 +/-0.42 | 1.664 +/-0.043 | 905 | Delrez et al. 2021 | Super-Ganymede |
| **TOI-776 b** | 8.25 | 4.00 +/- 0.9 | 1.85 +/-0.13 | 514 | Luque et al. 2021 | Gas-rich super-Earth |
| **TOI-1260 b** | 3.93 | 8.60 +/- 1.50 | 2.34 +/-0.11 | 860 | Georgieva et al. 2021 | Gas-rich super-Earth |
| **HD136352 d** | 107.25 | 8.82 +/-0.9 | 2.562 +.09/-.08 | 431 | Delrez et al. 2021 | Gas-Rich super-Ganymede |
| **Kepler 538 b** | 81.74 | 10.6 +2.5/-2.4 | 2.215 +.04/-.03 | 380 | Mayo et al. 2019 | Gas-rich super-Ganymede |
| **K2-263 b** | 50.8 | 14.8 +/-3.1 | 2.41 +/-0.12 | 470 | Mortier et al. 2018 | Gas-rich super-Earth |
| **Kepler 411 b** | 3.00 | 25.6 +/-2.6 | 2.401 +/-0.053 | 1138 | Sun et al. 2019 | Gas-rich super-Earth |
| **TOI-561 c** | 10.78 | 5.40 +/-0.98 | 2.878 +/-0.096 | 860 | Lacadelli et al. 2020 | Sub-Neptune |
| **Kepler 36 c** | 16.22 | 7.13 +/-0.18 | 3.679 +/-0.098 | 928 | Vissapragada et al. 2020 | Sub-Neptune |
| **GJ 1214 b** | 1.58 | 8.17 +/-0.43 | 2.742 +/-0.05 | 596 | Cloutier et al. 2021 | Sub-Neptune |
| **LTT 3780 c** | 12.25 | 8.60 +1.60/-1.30 | 2.30 +.16/-.15 | 353 | Cloutier et al. 2020 | Sub-Neptune |
| **TOI 431 d** | 12.46 | 9.90 +/-1.5 | 3.29 +/-0.09 | 633 | Osborn et al. 2021 | Sub-Neptune |
| **HD 136352 c** | 27.59 | 11.24 +/-0.06 | 2.916 +/-0.07 | 677 | Delrez et al. 2021 | Neptune |
| **K2-32 b** | 8.99 | 15.0 +1.8/-1.7 | 5.30 +/-0.19 | 837 | Lillo-Box et al. 2020b | Neptune |
| **GJ 436 b** | 2.64 | 23.1 +/-0.8 | 4.19 +/-0.109 | 686 | Turner et al. 2016 | Neptune |
| **K2-19 b** | 7.92 | 32.4 +/-1.7 | 7.0 +/-0.2 | 854 | Petigura et al. 2020 | Neptune |
| **TOI-849 b** | 0.77 | 39.09 +2.7/-2.6 | 3.44 +.16/-.12 | | Armstrong et al. 2020 | Neptune |
| **K2-108 b** | 4.73 | 59.4 +/-4.4 | 5.33 +/-0.21 | 1446 | Petigura et al. 2017 | Super-Neptune |
| **CoRot-8 b** | 6.21 | 69.3 +/-10.8 | 6.94 +/-0.2 | 870 | Raetz et al. 2019 | Super-Neptune |
| **WASP-127 b** | 3.38 | 52.35 +6.8/-5.5 | 14.69 +/-0.3 | 1400 | Seidel et al. 2020 | Saturn |
| **KOI-1783.01** | 134.46 | 71 +11/-9 | 8.86 +.25/-.24 | | Vissapragada et al. 2020 | Saturn |
| **NGTS-11 b** | 35.46 | 109 +29/-23 | 9.16+.31/-.36 | 435 | Gill et al. 2020 | Saturn |
| **Kepler 539 b** | 125.63 | 308 +/-92 | 8.37 +/-0.19 | 388 | Mancini et al. 2016 | Saturn |
| **TOI-1899 b** | 29.02 | 210 +/-22 | 12.9 +0.4/-0.6 | 362 | Cañas et al. 2020 | Jupiter |
| **CoRoT-9 b** | 95.27 | 267 +/-16 | 11.95 +0.8/-0.7 | 420 | Bonomo et al. 2017 | Jupiter |
| **Kepler 167 e** | 1071 | 321 +51/-48 | 10.16 +/-0.42 | 134 | Chachan et al. 2022 | Jupiter |
| **KOI-3680 b** | 141.24 | 613 +60/-67 | 11.1 +0.7/-0.8 | 347 | Hébrard et al. 2019 | Jupiter |
| **Kepler 1704 b** | 988.88 | 1319 +/-92 | 11.94 +.48/-.46 | 254 | Dalba et al. 2021 | Super-Jupiter |
| **Kepler 1514 b** | 217.83 | 1678 +/-70 | 12.42 +/-0.26 | 388 | Dalba et al. 2021 | Super-Jupiter |

The Solar System analog names for the Rock-Ice Giant class are "Neptune", "sub-Neptune", or "super-Neptune". Given the large impact from inflationary mechanisms on the radius of planets with massive H-He envelopes (e.g. Lopez & Fortney 2014; Zeng et al. 2019; Millholland et al. 2020), planetary mass provides a more reliable parameter for characterizing Rock-Ice Giants as "sub" or "super" Neptunes than planetary radius. For example, Kepler 87c is a Neptune composition planet (MPB20) with a super-Earth mass (6.4 +/- 0.8 $M_⊕$) and a sub-Saturn radius (6.14 +/- 0.29 $R_⊕$). Based upon the mass ranges for the composition classes identified in section 4, Kepler 87c falls into the mass range for Rock-Terrestrial planets (<10 $M_⊕$) and therefore can be classified as a "sub-Neptune" despite having a radius 50% larger than Neptune. As indicated in Table 4, the mass ranges for sub-Neptune, Neptune, and super-Neptune class planets are 5-10 $M_⊕$, 10-50 $M_⊕$, and 50-110 $M_⊕$ respectively. The sub-Neptune mass range overlaps with the Rock-Terrestrial population mass range (sections 4.1 and 4.3) whereas the super-Neptune mass range overlaps with the Gas Giant population mass range (section 4.4).

### ~5.3 Gas Giant Composition Classes

Gas Giants have an H-He envelope mass fraction >50% by mass with rock-ice (metal) fractions <50% by mass. Models of Jupiter and Saturn indicate metal fractions of <10% and 20-30% respectively (Helled 2018). Based upon exoplanet samples and formation mechanisms, the maximum mass of metals is expected to be in the range 50 – 100 $M_⊕$ for most Gas Giants (e.g. Fortney et al. 2007; Miller & Fortney 2011; Lambrechts et al. 2014; Mocquet et al. 2014) and therefore the metal fraction is expected to be less than 20% for most Jupiter mass and larger Gas Giants.

The Solar System analog names for Gas Giants are "Jupiter" for planets with <20% metals by mass and "Saturn" for planets with 20.0 – 49.9% metals by mass. The mass range of Saturn and Jupiter class planets potentially overlaps from 50 to 300 $M_⊕$. Millholland et al. (2020) found the H-He mass fraction of the super-Neptune K2-108b is only 5.9%. With a mass of 59.4 $M_⊕$, K2-108b would have a total metal mass of approximately 56 $M_⊕$ which indicates Saturn composition planets could reach at least 280 $M_⊕$. The upper mass limit for Saturn composition planets is therefore listed as 300 $M_⊕$ in Table 4. Kepler 539 b is a high mass Saturn composition planet with a mass 308 but a relatively large uncertainty of +/-92 $M_⊕$ (Mancini et al. 2016).

Based upon formation mechanisms, star forming gas collapse could form Brown Dwarfs as small as ~4 $M_{Jup}$ (Caballero 2018; Luhman & Hapich 2020) and disk instability could form Gas Giants as large as ~25 $M_{Jup}$ (Goda & Matsuo 2019). Following Lecavelier & Lissauer (2022), Brown Dwarfs are identified in Table 4 as Gas Giants with a mass exceeding the deuterium burning limit of approximately 13 $M_{Jup}$ regardless of formation mechanism and therefore will have metal mass fractions <10%. Note that if Brown Dwarfs are instead defined by formation mechanism (e.g. Chabrier et al. 2014), then the lower mass limit for the Brown dwarf composition class becomes 4 $M_{Jup}$ or ~1300 $M_⊕$ and would overlap the Jupiter composition class mass range from 4 – 25 $M_{Jup}$.

## 6 Discussion

### ~6.1 *Composition Classification as an Algorithm*

Stern & Levison (2002) suggested eight criteria for an algorithm to determine the classification of a body as a "planet". These criteria are that the definition: (1) be physically based, (2) be determined based upon observable characteristics, (3) be quantitative, (4) uniquely classify any given body, (5) be deterministic, (6) be comprised of the fewest possible criteria, (7) be robust to new discoveries, and (8) be reasonably backward compatible. While Stern & Levison (2002) proposed these criteria for an algorithm that would determine whether or not a body is a planet, it is worthwhile to note that the composition classification system described in this paper meets seven of these criteria.

The broad composition classes Rock-Terrestrial, Rock-Ice Terrestrial, gas-rich Terrestrial, Rock-Ice Giant, and Gas Giant are identified from the planetary mass fractions of H-He gas, rock and ice. This characteristic of the composition system (Table 1) meets five of the eight criteria proposed by Stern & Levison (2002). Specifically, the composition classification system is physically based, quantitative, uniquely classifies every planet's composition, is robust to new discoveries, and comprises a small number of classification criteria.

Note that while model degeneracy and data uncertainty will in some cases make it difficult to determine which of the classes a specific exoplanet should be assigned to, each planet can only have one of the broad composition classes. Therefore the composition classification system can uniquely classify every planetary body when the necessary data is available.

The system is also robust to future discoveries in several ways. First, since the broad composition classes are based upon the mass fractions of the H-He and rock-ice components, every newly discovered exoplanet must fall into one of the five broad composition classes. Second, the bulk composition classes described provide more detailed composition characterization. If new, exotic planetary compositions are discovered, these compositions can be added to the list of bulk compositions without abandoning those classes already described or requiring new broad composition classes. The carbon-enriched terrestrial composition provides an example of this flexibility as this bulk composition class represents an exotic composition not found in the Solar System. Third, since the composition classification system is not dependent upon interior structure models the system will not require constant revisions as equation of state and interior structure models improve (e.g. Helled & Stevenson 2017; Wahl et al. 2017; Hakim 2018b, 2019a).

The composition classification system also meets the criteria of being based upon observable characteristics as determination of the composition class of a planet is primarily derived from mass and radius measurements which are analyzed in the context of mass-radius-composition models that are derived from EoS models and the planet's stellar flux.

The composition system discussed in this paper is also "backward compatible" by which Stern & Levison meant to avoid numerous reclassifications that would confuse lay people. The five broad classes place Solar System planets into the same classes they have previously been assigned to with the slight modifications being: (1) that the traditional "Ice Giants" are now "Rock-Ice Giants" as explained in section 3.2 and (2) the Terrestrial planet class is divided into Rock-Terrestrial, Rock-Ice Terrestrial, and gas-rich Terrestrial. These slight modifications maintain familiar terminology while allowing the composition classification system to meet the condition of uniquely classifying any given body.

The only criterion from the Stern & Levison (2002) list that the composition classification system does not meet is that the system cannot be deterministic, by which they argued the classification of a body as a "planet" should not change over time. This criterion cannot fully apply to planetary composition due to the volatile nature of planetary envelopes which may be stripped through core powered mass loss, photo-evaporation, and giant impacts (Owen et al. 2020; Misener & Schlichting 2021; Gupta & Schlichting 2021; Ogihara et al. 2021). Therefore, planets that start out as Gas Giants or Rock-Ice Giants can evolve to a different composition class over time if the gas envelope is sufficiently stripped (e.g. Dai et al. 2019; Armstrong et al. 2020).

*~6.2 Usefulness of the Composition Classification System*

The composition classification system presented in this paper has a number of useful characteristics. The five broad composition classes ***Gas Giant***, ***Rock-Ice Giant***, ***gas-rich Terrestrial, Rock Terrestrial, and Rock-Ice Terrestrial*** provide familiar terminology originating with known Solar System composition classes. However, each composition class allows for a broader range of composition characteristics than found in the planetary population of the Solar System. All Solar System planets and all exoplanets may be grouped into one of these five classes depending only upon the mass fractions of H-He gas, rock, and ice.

Modeling the composition of exoplanets suffers from degeneracy and many planets will have a mass and radius consistent with several possible bulk compositions. For example, following Georgieva et al. (2021) TOI 1260b, $M_\oplus$ = 8.60 +1.40 -1.50, is a super-Earth mass Terrestrial planet that has a rock-ice Terrestrial composition with 50% Earth-like rocky core and 50% H₂O, or a gas-rich Terrestrial composition with an Earth-like core and approximately 0.1% by mass H-He envelope. The composition modeling for TOI 1260b highlights the usefulness of the composition classification scheme presented in this paper. Based upon the measured radius, TOI 1260b was identified as a "mini-Neptune" by Georgieva et al. (2021). However, the "mini-Neptune" characterization differs significantly from the compositional modeling which indicates the planet has a rock-ice Terrestrial or a gas-rich Terrestrial composition.

Another aspect of the composition classification scheme presented here is that Solar System analog names such as Mercury, Earth, Ganymede, Neptune, Saturn, Jupiter, and "sub" or "super" versions of these names can be consistently applied based upon composition modeling and planetary mass (Table 4). For example, TOI-1260b is a "super-Ganymede" if a rock-ice Terrestrial composition, or a "gas-rich super-Earth" if a gas-rich rock Terrestrial composition (Georgieva et al 2021).

Note that these Solar System analog names make sense when using the planetary mass in combination with the planetary composition models. The resulting characterization will better identify the characteristics of the planet than applying Solar System analog names based upon the planetary radius. The radius of TOI-1260b would classify it as a sub-Neptune, but the planet does not have a Rock-Ice Giant composition as the H-He fraction is much less than 1.0% by mass (Georgieva et al. 2021). With the classification system presented in this paper, Solar System analog names can be consistently used to describe planets based upon their mass and composition derived from M-R-composition models. This approach avoids the inconsistency between name and composition that results when using Solar System analog names based upon the planetary radius alone (e.g. TOI 1260b). As noted in the introduction, Kepler 87c is a super-Earth mass, sub-Saturn radius planet with a Neptune composition (MPB20). Applying the classification scheme presented in this paper, Kepler 87c has a Rock-Ice Giant composition and in combination with its mass can be characterized as a "sub-Neptune".

Planetary composition models also require knowledge of the stellar flux or equilibrium temperature in order to determine the mass fraction of H-He gas (Lopez & Fortney 2014; Zeng et al. 2019). The planet LTT 3780 c has both mass and radius values nearly identical to those of TOI-1260 b (Table 5). However, the equilibrium temperature of LTT 3780 c is 353 K (Cloutier et al. 2020) whereas the equilibrium temperature of TOI-1260 b is 860 K (Georgieva et al. 2021). As a result of the equilibrium temperature difference, both Lopez & Fortney (2014) and Zeng et al. (2019) models indicate that while TOI-1260 b could have approximately 0.1 percent by mass H-He and is a gas-rich super-Earth, LTT 3780 c has 1-2 percent by mass H-He and is a sub-Neptune. These two planets illustrate the need for equilibrium temperature values in addition to mass and radius data, to improve the determination of the composition class for planets with H-He envelopes.

It is important to note that the composition classification system describes composition characteristics only and not circumstances. For example, the system does not distinguish Rock Terrestrial planets with a solid surface from planets with a molten rock surface, such as lava worlds (Chao et al. 2021). Rock-Ice Terrestrial planets with a supercritical steam envelope are not distinguished from those with an ice surface and sub-surface oceans. These examples illustrate that additional terminology could be added to further describe the physical circumstances of planets.

Finally, the classification scheme can provide a means for consistently characterizing newly discovered exoplanets with names that correctly indicate the planetary composition when communicating those discoveries to students and the general public. For example, many planets characterized as "sub-Neptune" based upon the planetary radius are actually super-Ganymede, gas-rich super-Earth, or gas-rich super-Ganymede planets based upon the mass and composition analysis (e.g. Georgieva et al. 2021; Otegi et al. 2021).

## 7 Conclusion

This paper was motivated by the need, as discussed in the introduction, for a planetary composition classification system that allows for consistent characterization of currently known and newly discovered exoplanets. Current characterization of planetary composition in the literature contains significant variation in the usage of terminology and that terminology is sometimes inconsistent with the composition model of the planet.

The composition classification system described in this paper classifies every planet into one of five broad composition classes: Gas Giant, Rock-Ice Giant, gas-rich Terrestrial, Rock Terrestrial, or Rock-Ice Terrestrial. Planets are assigned one of these five composition classes based only upon the mass fractions of H-He gas and rock-ice (Table 1). Each broad composition class is subdivided into multiple possible bulk composition classes using Solar System analog names (Table 4). An important aspect of this classification system is that it covers all currently described composition classes but is flexible to future discoveries as any new or exotic composition classes could be added to the list of bulk composition classes, found in Table 4, without altering the broad and bulk composition classes already described. The composition classification system also does not rely on specific interior structure models and therefore will not require modification as equation of state and interior structure models are revised.

Empirical mass ranges for the composition classes were explored and in combination with the bulk composition classes provide useful guidelines for applying terms such as "super-Earth", "sub-Neptune", "super-Ganymede" and other Solar System analog descriptors (Tables 4 & 5).

The familiar terminology utilized for both the broad composition classes and the bulk composition classes, combined with the simplicity and flexibility of the system can facilitate consistent communication of Solar System and exoplanetary discoveries not only in the literature, but also to students and laypersons.

**Acknowledgements**

This research has made use of the NASA Exoplanet Archive, which is operated by the California Institute of Technology, under contract with the National Aeronautics and Space Administration under the Exoplanet Exploration Program. This research has made use of NASA's Astrophysics Data System Bibliographic Services. The author would like to thank Ravit Helled and Vardan Adibekyan for helpful comments and suggestions on earlier drafts of this paper.